\documentclass[review,12pt]{elsarticle}

\usepackage{amssymb}
\usepackage{amsmath}

\journal{Signal Processing}

\begin{document}

\begin{frontmatter}



\title{Pulse Processing --- Overview and Challenges}


\author{Jonathan H. Manton} 

\affiliation{organization={Department of Electrical and Electronic Engineering},
            addressline={The University of Melbourne}, 
            city={Parkville},
            postcode={3010}, 
            state={Victoria},
            country={Australia}}

\begin{abstract}
The detection of irregularly spaced pulses of non-negligible width is a fascinating yet under-explored topic in signal processing. It sits adjacent to other core topics such as radar and symbol detection yet has its own distinctive challenges. Even modern techniques such as compressed sensing perform worse than may be expected on pulse processing problems. Real-world applications include nuclear spectroscopy, flow cytometry, seismic signal processing and neural spike sorting, and these in turn have applications to environmental radiation monitoring, surveying, diagnostic medicine, industrial imaging, biomedical imaging, top-down proteomics, and security screening, to name just a few. This overview paper endeavours to position the pulse processing problem in the context of signal processing. It also describes some current challenges in the field.
\end{abstract}

\end{frontmatter}



\section{Introduction}
\label{sec:intro}

\begin{figure}[t]
    \centering
    \includegraphics*[width=\textwidth]{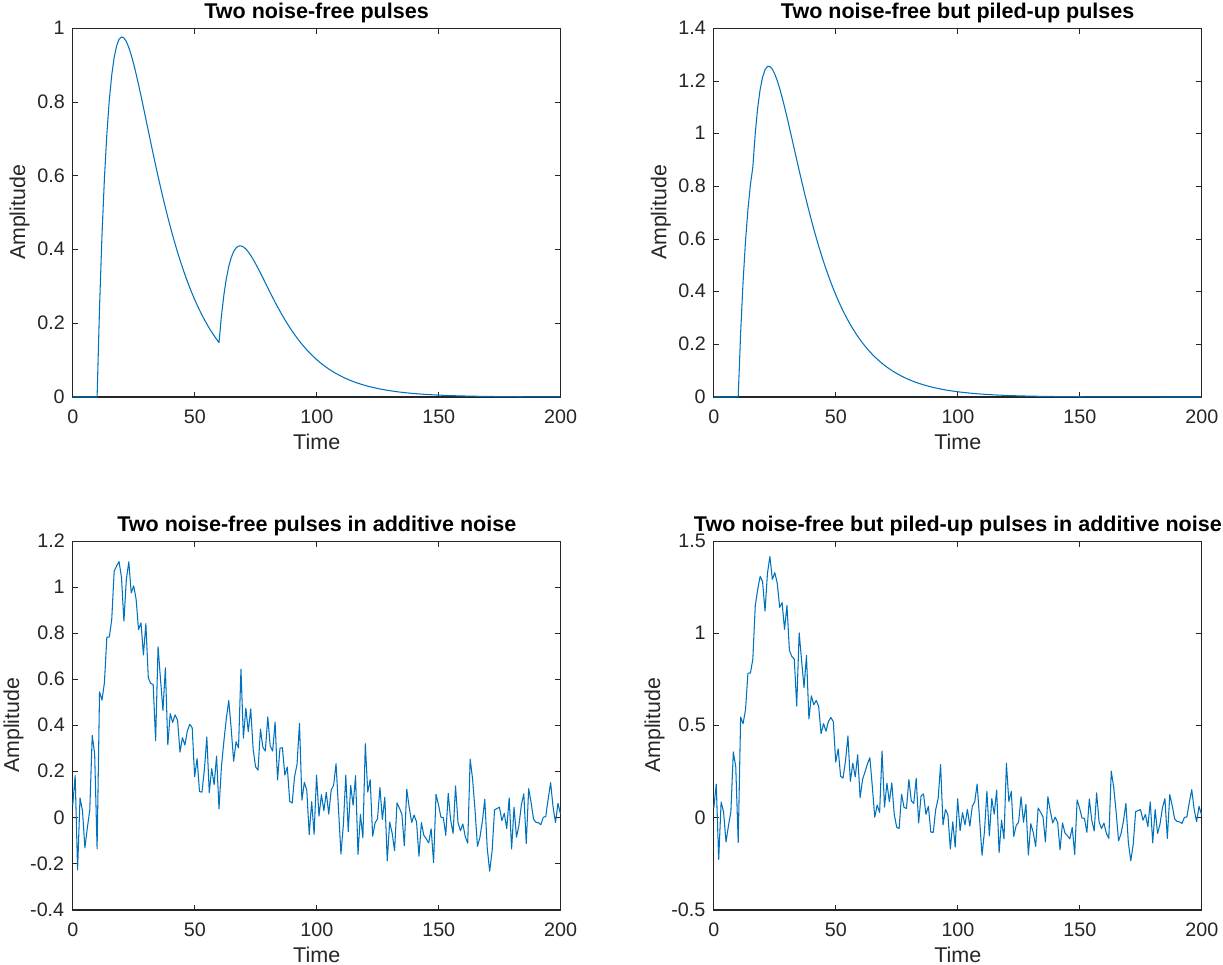}
    \caption{The top left figure contains two pulses of the same shape but different amplitudes. One pulse starts at time 10 and the other at time 60. In the absence of noise the two pulses are readily identifiable. Once noise is added, as in the bottom left figure, it becomes harder to determine the number and locations of the two pulses. Besides noise, the other main challenge is when two pulses occur almost on top of each other, as in the top right figure, where one pulse starts at time 10 and the other at time 16. As soon as noise is added to this, as in the bottom right figure, it becomes exceedingly difficult to determine the number and locations of piled-up pulses.}\label{fig:pp}
\end{figure}

Pulse processing\footnote{The reader is cautioned of another usage of ``pulse signal processing'' which involves pulses of essentially zero width~\cite{nallathambiTheoryAlgorithmsPulse2020}.} refers to extracting information from a stream of pulses
\begin{equation} \label{eq:yt}
    y(t) = \sum_i \alpha_i p(t - \tau_i) + w(t)
\end{equation}
where $p(t)$ is a known pulse. Here, $\sum_i \alpha_i p(t - \tau_i)$ represents a sum of pulses of differing amplitudes $\alpha_i$ and arrival times $\tau_i$. The pulse $p(t)$ is assumed to have non-negligible width, meaning it is possible two or more pulses will overlap, or ``pile up'' on each other (Figure~\ref{fig:pp}). The archetypal pulse processing problem is recovering the amplitudes and arrival times given the noisy observations $y(t)$, where $w(t)$ in \eqref{eq:yt} is additive white Gaussian noise. While posed in continuous time, usually $y(t)$ is sampled uniformly in time, and indeed, throughout the paper, the reader may generally assume $t$ is an integer. In practice, $y(t)$ is only observed over a finite interval, and the number of pulses is unknown.

\begin{figure}[t]
    \centering
    \includegraphics*[width=\textwidth]{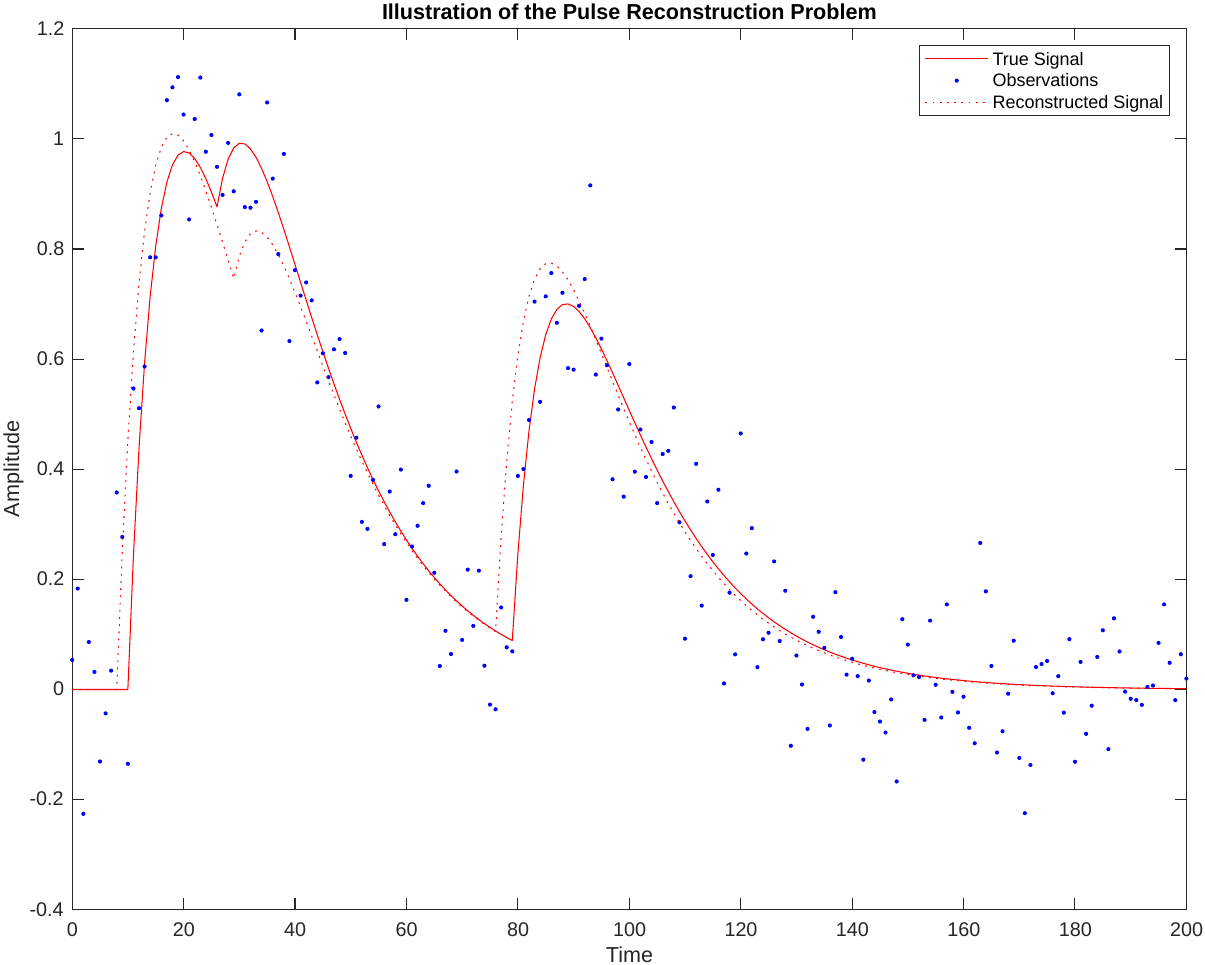}
    \caption{Shown in solid red are three super-imposed pulses. Sampling this true signal at integer times and then adding noise results in the noisy observations shown as blue dots. Given these observations, the basic aim of pulse processing is identifying each individual pulse and thus reconstructing the original signal. An illustrative example of a reconstruction is the dotted red curve. Note the inevitable errors in the pulse arrival times and pulse amplitudes.}
    \label{fig:pr}
\end{figure}

For concreteness, an illustrative pulse reconstruction problem is shown in Figure~\ref{fig:pr}. The true signal is the sum of three pulses, each pulse having the same shape but different amplitude and arrival time. From the noisy observations, the challenge is to identify each and every pulse. In this hypothetical example, the reconstruction correctly determined there were three pulses, but made errors in estimating the pulse amplitudes and arrival times.

The majority of papers involving the model \eqref{eq:yt} are focused on developing algorithms for specific applications. An early exception is a 1980 paper that studied the problem theoretically~\cite{kwakernaakEstimationPulseHeights1980}; its opening paragraph stated that this problem ``seems to have been less extensively studied'' than the simpler problem of finding just a single pulse. We share this sentiment; while identifying pulses feels like a fundamental signal processing problem, it has not been as thoroughly studied as other fundamental problems.

The pulse $p(t)$ in \eqref{eq:yt} could model the action potential of a neuron, or the response of a scintillation detector to an incoming photon, to give just two examples. An approximate model of the latter is a double-exponential pulse
\begin{equation} \label{eq:pt}
    p(t) = \begin{cases}
        0,\quad &t < 0, \\
        \exp(-a t) - \exp(-bt),\quad &t \geq 0
    \end{cases}
\end{equation}
where the known coefficients $a>0$ and $b>0$ control the rise and decay times. Figure~\ref{fig:pp} shows examples of two such pulses added together, with $a=0.06$ and $b=0.15$.

Pulse processing becomes challenging when pulses pile up. Pile-up refers to two closely-spaced pulses $\alpha_1 p(t-\tau_1) + \alpha_2 p(t-\tau_2)$ where $|\tau_1-\tau_2|$ is small compared to the ``width'' of a single pulse, as illustrated in the top right of Figure~\ref{fig:pp}. Piled-up pulses can become indistinguishable from a single pulse once noise is added. The bottom plots in Figure~\ref{fig:pp}, which are noisy versions of the top plots, each have a rising edge that is not everywhere increasing, so it is hard to know whether the rising edge comprises one, two or many individual pulses.

An obvious generalisation of the pulse processing problem is allowing for variations in the pulse shape by replacing $p(t-\tau_i)$ with $p_i(t-\tau_i)$ where the $p_i$ are restricted to a family of known pulses. This can model the changes in a photon detector's output caused by differences in the trajectory of an incoming photon. It also has applications to neural spike sorting~\cite{lewickiReviewMethodsSpike1998,pachitariuFastAccurateSpike2016}. Another generalisation is to higher dimensions, with applications to astronomy and low-intensity imaging~\cite{sequeiraImageDetectionLowlevel1993}.

\begin{figure}[t]
    \centering
    \includegraphics*[width=\textwidth]{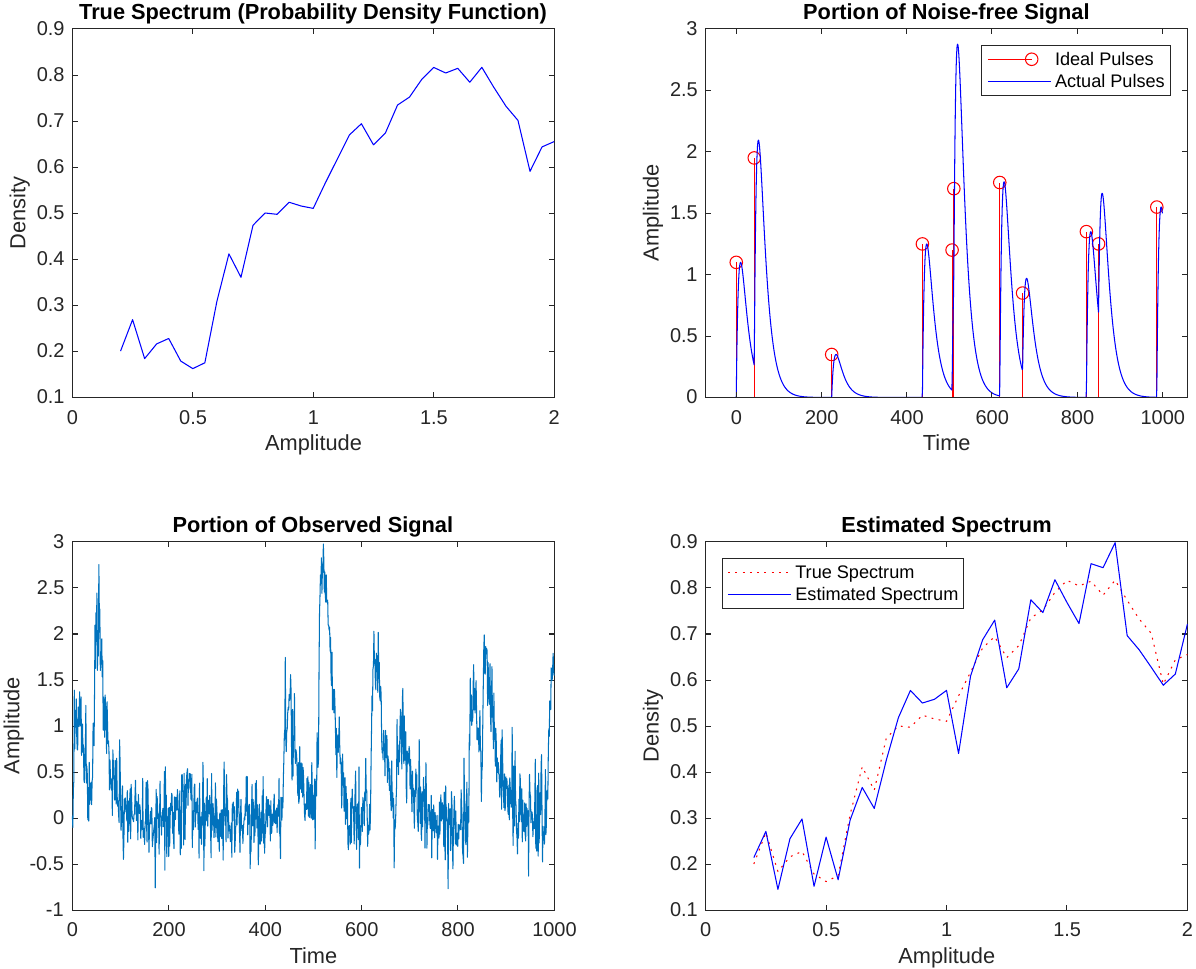}
    \caption{The top-left figure shows a hypothetical amplitude spectrum. It represents the probability density function from which the heights of the ideal pulses, shown in red in the top-right figure, are drawn independently. The arrival times of the ideal pulses are generated by a homogeneous Poisson process. The actual pulses, shown in blue, are obtained by convolving the ideal (Dirac-delta) pulses $\alpha_i \delta(t - \tau_i)$ with the pulse shape $p(t)$. Equivalently, each pulse is of the form $\alpha_i p(t - \tau_i)$, and all the pulses are added together. Adding noise to this yields the observed signal shown in the bottom left figure. In spectroscopy, the analyser takes this observed signal and produces an estimate of the amplitude spectrum. The most common way of doing this is by first identifying individual pulses in the observed signal, determining their amplitudes, then plotting a histogram of these amplitudes. A possible output of the spectrometer is shown as the blue trace in the bottom right figure. The dotted red trace is the same spectrum as in the top left figure.}
    \label{fig:spec}
\end{figure}

In the other direction, the pulse processing problem can be made more specific by assuming the amplitudes $\alpha_i$ and arrival times $\tau_i$ in \eqref{eq:yt} are stochastic processes. Nuclear spectroscopy~\cite{knollRadiationDetectionMeasurement2010} corresponds to the particular situation where the arrival times are Poisson and the amplitudes $\alpha_i$ are independent and identically distributed samples from an unknown distribution $\mathcal{A}$. In some applications, it suffices to estimate the so-called spectrum $\mathcal{A}$, which is potentially easier than identifying each and every pulse. This is illustrated in Figure~\ref{fig:spec} and will be taken up in \S\ref{sec:spec_est}.

Hardware implementations of pulse processing algorithms remain of considerable interest. Recent advances in high-speed analogue-to-digital converters (ADC) allow analogue preprocessing techniques to be replaced by digital computation by placing the ADC closer to the raw output of the detector~\cite{abbeneRealTimeDigital2013, difulvioDigitalPulseProcessing2016}. The high data rates, often greater than 100 MHz sampling~\cite{asztalosGeneralPurposePulse2016}, produce vast amounts of data. In this environment, FPGAs have proven suitable for online pulse processing~\cite{leeDevelopmentFPGAbasedDigital2013,saxenaDigitalPulseDeconvolution2020a}. Different trade-offs lead to different solutions, such as high-throughput solutions~\cite{yangHighThroughputRate2020} or low-cost solutions~\cite{caoFlexibleLowcostFPGAbased2021}.

The remainder of the paper is organised as follows. The next section anticipates readers are already drawing connections between pulse processing and topics they are more familiar with. By discussing these neighbouring topics up front, the reader can then focus freely on the foundations of pulse processing presented in \S\ref{sec:foundations}.
There, the underlying problem of interest is identifying the individual pulses in \eqref{eq:yt}. In some applications though, the end goal is to estimate the amplitude distribution of the pulses. This is discussed in \S\ref{sec:spec_est} where it is explained how the amplitude distribution can be estimated without finding the individual pulses. Finally, \S\ref{sec:conclusion} concludes the paper. It lists a few open challenges in pulse processing.

\section{Neighbouring Problems}

When encountering the pulse processing problem for the first time, a number of similar problems come to mind. The following subsections highlight the similarities with, and differences between, these neighbouring problems.

\subsection{Matched Filter} \label{sec:matched_filter}

The matched filter is often the first tool that comes to mind because it is the optimal detector for deciding whether a signal is present. A minor extension allows it to be applied when the signal to be detected may have been shifted in time by an unknown amount~\cite{ehrenbergSignalprocessingTechniquesResolving1978}.

Precisely, if we wish to determine whether $y(t)$ contains zero pulses or one pulse, we can convolve it with the time-reversed pulse $p(-t)$; this is the matched filter. If $y(t)=p(t-\tau) + w(t)$ then the output of the matched filter is
\begin{equation}
    p(-t) \ast y(t) = p(-t) \ast p(t-\tau) + p(-t) \ast w(t)
\end{equation}
where $\ast$ denotes convolution. By the Cauchy-Schwarz inequality,
\begin{equation}
    p(-t) \ast p(t-\tau) = \int p(t'-t) p(t'-\tau)\,dt'
\end{equation}
takes its largest value when $t=\tau$. Therefore, a standard approach for detecting a single pulse is to find the time $t$ at which the output of the matched filter $p(-t) \ast y(t)$ is maximised, and provided the value of the output at this time is above a certain threshold, declare that a pulse has been found.

Despite this result, the matched filter does not solve the pulse processing problem because it is not designed to handle pulse pile-up~\cite{ehrenbergSignalprocessingTechniquesResolving1978}. Intuitively at least, pile-up would be easier to handle if the pulses are made narrower, and there is no reason to expect the matched filter will produce narrower pulses. Building on this comes the idea of pulse shaping, whereby the matched filter is replaced by a pulse-shaping filter, with the same aim of looking for peaks in the filtered output; see~\S\ref{sec:pulse_shaping} for details.

\subsection{Radar}

Radar works by transmitting a pulse $p(t)$ towards an object then waiting for a return signal $y(t) = \alpha\,p(t-\tau) + w(t)$ where $w(t)$ is noise~\cite{richardsFundamentalsRadarSignal2022}. Determining $\tau$ gives an estimate of how far away the illuminated object is. If there are multiple objects (or clutter) then the return signal will contain multiple pulses $\alpha_i p(t-\tau_i)$, bringing it closer to the form \eqref{eq:yt}. There are a number of key differences to be aware of though.

Implicit in \eqref{eq:yt} is that there are an infinite number of pulses. Even though in practice we only ever work with finite data, the pulse processing problem is especially challenging when there are a very large number of pulses to contend with. In radar, there are generally only a small number of pulses of interest (and the rest can be treated as clutter). Moreover, in radar, multiple pulses are sent over time and hence slow-moving objects can be tracked, making them easier to detect in noise.

In active radar systems, the pulse shape can be engineered to facilitate subsequent pulse processing. (The mathematics behind this involves the radar ambiguity function~\cite{moranGroupTheoryRadar2004}.) This luxury is not available in other applications where the pulse shape cannot be modified freely.

\subsection{Symbol Detection in Communication Systems}

Communication systems often transmit data digitally symbol by symbol~\cite{molischWirelessCommunicationsFundamentals2023}. Each symbol can be considered a pulse and therefore the transmitted signal is a pulse train, such as $\sum_i \alpha_i\,p(t-i)$ for binary phase-shift keying, where each $\alpha_i$ is $\pm 1$. In multipath environments, where there is more than one path between transmitter and receiver, each symbol $\alpha_i\,p(t-i)$ might be received as a finite sum of pulses $\sum_j \alpha_i \beta_j \, p(t-i-\tau_j)$ where $\beta_j$ and $\tau_j$ are the attenuation and total delay of the $j$th path.

While the received signal will be of the form \eqref{eq:yt}, the communications problem has a specific structure that makes its various solutions inapplicable to the more general pulse processing problem. Without multipath, the time-of-arrival of all pulses are known. Even with multipath, normally the small number of delays $\tau_j$ will vary slowly and hence can be estimated over a period of time relatively easily. This is especially true since the symbol alphabet (the set of possible values of the $\alpha_i$) is known to the receiver.

\subsection{Nonlinear Regression} \label{sec:nlr}

Given a finite interval $[a,b]$ we may seek to fit a finite number $N$ of pulses to $y(t)$ for $t \in [a,b]$ by minimising the mean-square residual
\begin{equation} \label{eq:msr}
    \int_a^b \left| y(t) - \sum_{j=1}^N \hat \alpha_j \, p(t - \hat \tau_j) \right|^2 \, dt
\end{equation}
or its discrete-time sampled equivalent.
This is a nonlinear regression problem and a nonlinear least-squares problem. It also corresponds with the maximum likelihood estimate (MLE) if the noise is white Gaussian and if we ignore end effects. If there is frequent pile-up then it may not be satisfactory to ignore end effects: pulses from before time $a$ added together might have a significant effect on $y(t)$ for $t \geq a$. There is also the issue of how to choose $N$.

Nevertheless, this is a very valid approach to the pulse processing problem, and the number of pulses $N$ can be estimated either by a model-order selection principle~\cite{kwakernaakEstimationPulseHeights1980} or, if the noise variance is known, then by choosing the $N$ that leaves a mean-square residual closest to its expected value~\cite{ehrenbergSignalprocessingTechniquesResolving1978}. More about the MLE approach will be said in \S\ref{sec:mle}.

Nonlinear regression and MLE approaches are not the be-all and end-all because the computations grow rapidly with $N$ and soon become intractable. Indeed, minimising \eqref{eq:msr} is an $N$-dimensional nonconvex optimisation problem whose global minimum is difficult to find. (Note \eqref{eq:msr} is quadratic in the $\hat\alpha_j$ and thus there is a closed-form expression for the best $\hat\alpha_j$ given the $\hat\tau_j$.)

\subsection{Frequency Domain} \label{sec:frequency_domain}

\begin{figure}[t]
    \centering
    \includegraphics*[width=\textwidth]{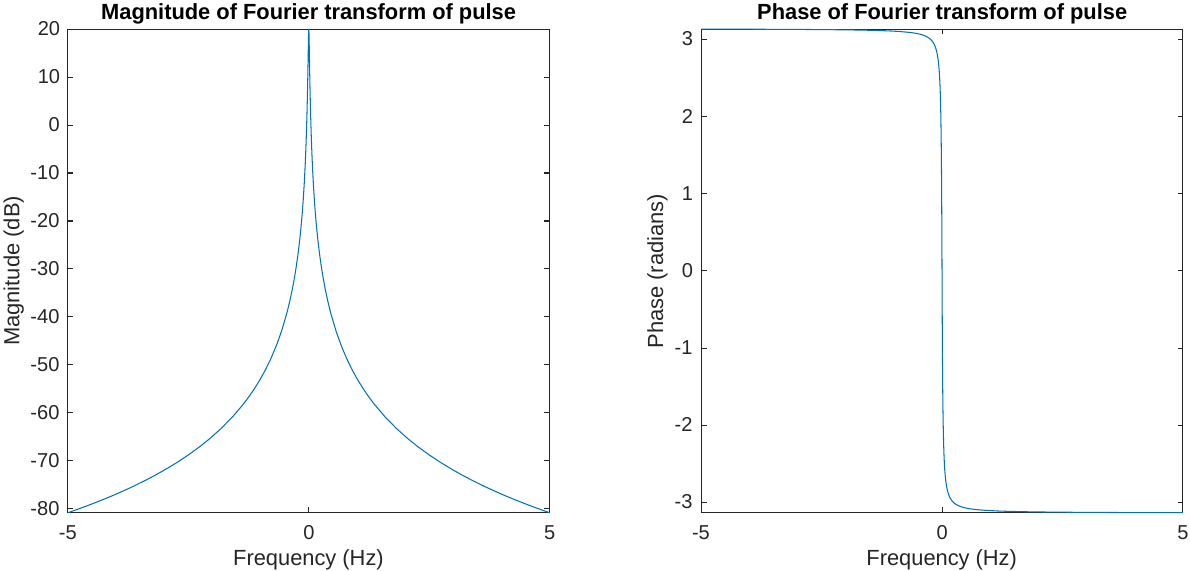}
    \caption{Graphs showing the Fourier transform of the doubly exponential pulse $p(t) = e^{-0.06t}-e^{-0.15t}$.}\label{fig:fd}
\end{figure}

The Fourier transform of the pulse stream \eqref{eq:yt} is
\begin{equation} \label{eq:Yf}
    Y(f) = \left( \sum_i \alpha_i e^{-\jmath 2 \pi \tau_i f} \right) P(f) + W(f)
\end{equation}
where $P(f)$ and $W(f)$ are the Fourier transforms of the pulse $p(t)$ and noise $w(t)$ respectively. The Fourier transform of the doubly exponential pulse \eqref{eq:pt} is
\begin{equation}
    P(f) = \frac1{a+\jmath 2 \pi f} - \frac1{b + \jmath 2 \pi f}.
\end{equation}
Its magnitude and phase are plotted in Figure~\ref{fig:fd} for the values $a=0.06$ and $b=0.15$. If we reduce the ``width'' of this pulse from 100 seconds to 1 $\mu$s then the ``frequency content'' increases from 1 Hz to 100 MHz. Put another way, as a rule of thumb, we would want to sample fast enough that a typical pulse lasts for at least 100 samples.

Despite the simple form of \eqref{eq:Yf}, there do not appear to be any significant benefits from working in the frequency domain instead of the time domain. Being a linear transformation, the Fourier transform will not convert the nonconvex problem of minimising the mean-square residue \eqref{eq:msr} into a convex problem, for example.

\subsection{Stochastic Differential Equations with Jumps}

For certain pulse shapes, the model \eqref{eq:yt} can be rewritten as a stochastic differential equation with jumps, where the jumps occur at the pulse arrival times. Take for example the model in~\cite{melansonDatadrivenInferenceStationary2019}: $dY(t) = F(Y(t)) \, dt + \sqrt{2D} \,dW(t) + dJ(t)$. A drift function $F$ of the form $F(Y(t)) = \alpha Y(t)$, where $0 < \alpha < 1$, will capture an exponentially decaying pulse. The noise term $\sqrt{2D}\,dW(t)$, where $W(t)$ is Brownian motion, models ``additive white Gaussian noise'' with noise intensity $D$. The compound Poisson process $J(t) = \sum_{i=0}^{N_\lambda(t)} B_i$ models pulse arrivals: $N_\lambda(t)$ is a Poisson process with rate $\lambda$, and the $B_i$'s are independent and identically distributed jump amplitudes.

Modelling using stochastic differential equations with jumps has been used in finance~\cite{leeJumpsFinancialMarkets2005} and neuroscience~\cite{melansonDatadrivenInferenceStationary2019}. In principle, ideas from stochastic differential equations with jumps can be applied to the pulse processing problem. However, this is a very broad class of models, and in the common situation where the pulse arrival times are Poisson distributed, the model \eqref{eq:yt} belongs to the more specific class of models known as filtered Poisson processes, albeit in additive noise, details of which will be provided in~\S\ref{sec:fpp}.

\section{Foundations and Approaches}
\label{sec:foundations}

The model \eqref{eq:yt} leads to different problem formulations depending on whether the $\alpha_i$ and $\tau_i$ come from a statistical distribution, and whether estimating the $\alpha_i$ and $\tau_i$ is the primary objective. Three common formulations are now described.

\paragraph*{No Statistical Information}
If nothing is known about the $\alpha_i$ and $\tau_i$ then their estimation from \eqref{eq:yt} is a parameter estimation problem with the added complication that the number of parameters is unknown. A starting point for an ``optimal'' solution is the maximum-likelihood estimator (\S\ref{sec:mle}).

\paragraph*{Stochastic Pulse Heights and Arrival Times}
A common assumption is that the arrival times $\tau_i$ are Poisson distributed with fixed rate $\lambda$, and the amplitudes $\alpha_i$ are drawn independently from a fixed distribution $\mathcal A$. This makes \eqref{eq:yt} a filtered Poisson process in the noise-free case (\S\ref{sec:fpp}). Usually $\mathcal A$ is not known.

\paragraph*{Estimation of Amplitude Distribution}
With the $\tau_i$ Poisson distributed and the $\alpha_i$ drawn independently from $\mathcal A$, sometimes it suffices to estimate only the amplitude distribution $\mathcal A$. While the majority of methods use a histogram of the estimated amplitudes as an estimate of $\mathcal A$, it is possible, at least in theory, to estimate $\mathcal A$ directly. This will be discussed further in \S\ref{sec:spec_est}. Another advantage of this formulation is that taking the histogram of only non-piled-up pulses will still serve as a sensible estimate of $\mathcal A$ (\S\ref{sec:pile_up_rejection}).

It is remarked that \eqref{eq:yt} can be interpreted as a filtered version of a Dirac impulse train. To wit, \eqref{eq:yt} is equivalent to 
\begin{equation}
    y(t) = \left( \sum_i \alpha_i \, \delta(t-\tau_i) \right) \ast p(t) + w(t)
\end{equation}
and the term in the brackets corresponds to a spike train with Dirac impulses occurring at times $\tau_i$ with corresponding amplitudes $\alpha_i$. Convolution with $p(t)$ is equivalent to applying a filter with impulse response $p(t)$.

\subsection{Maximum-Likelihood Estimation} \label{sec:mle}

With the noise $w(t)$ in \eqref{eq:yt} being white Gaussian, the maximum-likelihood estimates (MLEs) of the amplitudes $\alpha_i$ and arrival times $\tau_i$ are found by minimising the nonlinear least-squares problem \eqref{eq:msr}.
This is appropriate if no statistical information is available about the parameters $\alpha_i$ and $\tau_i$. There is a subtlety: the number of pulses $N$ in the interval $[a,b]$ of interest is generally not known. Therefore, the MLE is computed for different choices of $N$, and subsequently the ``best fit'' is chosen~\cite{kwakernaakEstimationPulseHeights1980,ehrenbergSignalprocessingTechniquesResolving1978}. What is meant by best fit is considered in the model-order selection literature~\cite{stoicaModelorderSelectionReview2004}.

As mentioned in \S\ref{sec:nlr}, \eqref{eq:msr} is quadratic in the $\alpha_i$ and hence there exists a closed-form expression for the $\alpha_i$ in terms of the $\tau_i$. Nevertheless, fitting $N$ pulses requires solving an $N$-dimensional nonconvex optimisation problem over the variables $\tau_1 < \tau_2 < \cdots < \tau_N$. Each $\tau_i$ lies in the compact interval $[a,b]$, hence the smaller $b-a$ is, the better. Finding the optimal solution is difficult and computationally intensive even for $N$ small. (We are optimistic that the theory in~\cite{mantonOptimisationGeometry2013} can be used to find the global optimum in finite time, but it is unlikely to be computationally practical for large $N$.)

Detecting pulses with small amplitudes is problematic. This can be rigorously shown by applying standard detection theory (\S\ref{sec:matched_filter}) to the simpler problem of detecting the presence of a single pulse. The smaller the energy $\int |\alpha\,p(t)|^2\,dt$ relative to the noise, the harder it is to detect the pulse $\alpha\,p(t)$.

Another intrinsic difficulty is pulse pile-up. It is demonstrated in~\cite{mcleanLimitationsDecisionBased2018} that as the distance $\tau_{i+1}-\tau_i$ between two smooth pulses becomes smaller, it becomes harder to distinguish whether there is a single pulse or two pulses, becoming impossible as $\tau_{i+1}-\tau_i \rightarrow 0$. (While the doubly exponential pulse \eqref{eq:pt} is not smooth at the origin, it is arbitrarily close to a smooth pulse obtained by convolving it with a narrow Gaussian distribution.)

As is to be expected, the MLE is sensitive to errors in the pulse shape. If the MLE expects pulses to be of the form $\tilde p(t)$ yet is applied to data generated using pulses $p(t)$ then the resulting errors in the $\alpha_i$ and $\tau_i$ can be large~\cite{mohammadian-behbahaniComparisonStudyPilecorrection2020}.

\subsection{Marked, Compound and Filtered Poisson Processes}
\label{sec:fpp}

When the arrival times are Poisson distributed and the pulse amplitudes are drawn independently from a fixed distribution, the model \eqref{eq:yt} is a noisy filtered Poisson process. Three related definitions are first given; see~\cite{snyderRandomPointProcesses1991} for further details.

A \emph{marked Poisson process} is a Poisson process with points $\tau_i$ that has associated with each point $\tau_i$ an additional variable $\alpha_i$ called a mark.

A \emph{compound Poisson process} is a special type of marked Poisson process whereby the marks $\alpha_i$ are mutually independent and identically distributed random variables that are also independent of all the $\tau_j$.

A \emph{filtered Poisson process} is a process of the form
\begin{equation} \label{eq:fpp}
    y(t) = \begin{cases}
        0,\quad &N(t)=0 \\
        \sum_{n=1}^{N(t)} h(t, \tau_n, \alpha_n),\quad &N(t) \geq 1,
    \end{cases}
\end{equation}
where the $\tau_i$ and $\alpha_i$ are a compound Poisson process and $N(t)$ denotes the number of points (pulses) up to and including time $t$. By introducing the counting process $N(t)$, it is implicitly being assumed that there is a starting time before which no pulses have arrived. For convenience this starting time is taken to be zero. Thus, $\tau_i$ is defined for $i \geq 1$ and satisfies $0 \leq \tau_1 < \tau_2 < \cdots$.

A \emph{noisy filtered Poisson process} (not defined in~\cite{snyderRandomPointProcesses1991}) is a filtered Poisson process corrupted by additive noise. We will implicitly assume throughout that this noise is white Gaussian.

Comparing \eqref{eq:fpp} with \eqref{eq:yt} shows that the latter is a noisy filtered Poisson process with the choice $h(t, \tau_n, \alpha_n) = \alpha_n\,p(t - \tau_n)$. This choice is used to model atmospheric noise in~\cite[Example 5.2.4]{snyderRandomPointProcesses1991}.

Noisy filtered Poisson processes have been discussed in the physics literature~\cite{theodorsenStatisticalPropertiesFiltered2017a}. They have been used earlier in the signal processing literature for related problems. In the simpler situation where $\alpha_i = 1$ for all $i$, a nonlinear filter was derived in~\cite{komaeeNonlinearFiltersBayesian2011} for estimating the arrival time of a single pulse. This was extended to multiple pulses in the low-intensity regime where pile-up is unlikely to occur~\cite{komaeeEstimationLowIntensityFiltered2012}.

Other works have considered the situation where the Poisson intensity $\lambda$ is allowed to vary with time. If $\lambda(t)$ is known except for an unknown delay, MAP and MMSE estimators for the unknown delay were derived in~\cite{heroTimingEstimationFiltered1991} for the special case of $\alpha_i = 1$ for all $i$. This was extended in~\cite{antoniadisTimedelayEstimationFiltered1994} to when the $\alpha_i$ are Gaussian. A modified EM algorithm for finding the MLE was derived. The case where the $\alpha_i$ have a geometric distribution was considered recently in~\cite{merhavOptimalCorrelatorsDetection2021}. This geometric distribution models the random avalanche gains of an avalanche photo-diode. There, an optimal correlator was derived for estimating the time delay.

In general, estimating pulse amplitudes and arrival times from a realisation of a noisy filtered Poisson process suffers from the same difficulties as the MLE approach in \S\ref{sec:mle}: high computational complexity, made worse by not knowing the number of pulses in any given region.

Numerical Bayesian approaches are broadly applicable to nonlinear estimation problems. The specific family of 
reversible jump Markov chain Monte Carlo methods~\cite{greenReversibleJumpMarkov1995, hastieModelChoiceUsing2012} can handle the model-order selection problem of not knowing the number of pulses. An implementation for noisy filtered Poisson processes has been given in~\cite{andrieuBayesianDeconvolutionNoisy2001}.

\subsection{Pulse Shaping} \label{sec:pulse_shaping}

Approaches based on the likelihood function, such as those discussed in the previous two sub-sections, are computationally intensive if not intractable. Rather than starting with a theoretically optimal solution then attempting to approximate it to reduce computational complexity, an alternative is to start with the optimal solution for a simpler problem then heuristically modify it to handle more complicated problems. This leads to the prevalent idea of pulse shaping whereby the matched filter, which is optimal for detecting a single pulse (\S\ref{sec:matched_filter}), is replaced by a different filter aimed at achieving a balance between finding each pulse accurately while mitigating the problems of pulse pile-up.

The basic idea is to apply a pulse shaping filter to the observations $y(t)$ in \eqref{eq:yt}. If $h(t)$ is the impulse response of the filter then the filter output is the convolution $h(t) \ast y(t)$. The peaks of this filtered signal are then taken to be the locations of the original pulses. Originally, in spectroscopy applications, the pulse shaping filter was designed so that the amplitude of each peak would be proportional to the energy of the corresponding incoming photon; see~\cite{konradDetectorPulseShaping1968} and the references in~\cite{qinPulseshapeDiscriminationMethod2018}. Alternatively, more sophisticated pulse-detection methods can be applied to this filtered signal.

A naive idea is to use the ideal inverse filter to convert each pulse to a Dirac impulse~\cite{ehrenbergSignalprocessingTechniquesResolving1978}. This filter might be ideal under perfect noise-free conditions but it is not a practical solution because it amplifies the noise unacceptably. (This can be seen by working in the frequency domain (\S\ref{sec:frequency_domain}) where the filtering operation becomes multiplication by $1/P(f)$. At frequencies where $P(f) < 1$, the noise is amplified.) Instead, one can design a filter which minimises the mean-square error between the actual output and the desired impulsive output given a noisy input~\cite{ehrenbergSignalprocessingTechniquesResolving1978}; this is the same as Wiener deconvolution. Note though that this optimality does not translate into optimal estimates of pulse location and amplitude.

The optimal pulse shaping filter is a trade-off~\cite{jordanovDigitalSynthesisPulse1994}.
Traditionally, a shaper that transforms an exponentially decaying pulse into a trapezoid is often used~\cite{radekaTrapezoidalFilteringSignals1972,jordanovDigitalSynthesisPulse1994}. Over a limited range at least, the rise time of the pulse does not affect the peak height of the output trapezoid; the peak height is proportional to the area under the original pulse. More recently, transforming pulses to cusp shapes~\cite{jordanovExponentialSignalSynthesis2012a} and Mexican hat wavelet pulses~\cite{qinPulseshapeDiscriminationMethod2018} have been considered, and even more recently, doubly exponential pulses have been transformed to Gaussian pulses~\cite{wangGaussianShaperNuclear2022}.

An extension of this idea is to use two or more pulse shapers. A filter producing a short pulse can be used to determine whether pile-up is present (in which case the pulse is rejected; see \S\ref{sec:pile_up_rejection}) while a trapezoidal filter is subsequently used to estimate pulse height~\cite{leeDevelopmentFPGAbasedDigital2013}.
Similarly, different pulse shaping filters can be used for detecting arrival times versus amplitudes~\cite{abbeneRealTimeDigital2013}. Adaptive pulse shaping has also been considered~\cite{saxenaDigitalPulseDeconvolution2020a}.

\subsection{Pile-up Rejection} \label{sec:pile_up_rejection}

Pile-up makes optimal methods intractable. Without pile-up, each pulse can be fitted one-by-one. In some applications it suffices to ignore significantly piled-up pulses and simply focus on identifying the arrival times and amplitudes of pulses that are not badly piled up. An example is spectroscopy, where the aim is to estimate the distribution of the amplitudes $\alpha_i$. Ignoring piled-up pulses does not change the distribution. The trade-off is having to wait longer to collect enough non-rejected pulses to estimate the distribution reliably.

A basic pile-up rejection method is to reject a pulse if its width is too large~\cite{abbeneRealTimeDigital2013}. Like all methods though~\cite{mcleanLimitationsDecisionBased2018}, pile-up will rarely be detected if a pulse occurs on the rising edge of another pulse~\cite{yangHighThroughputRate2020}.

\subsection{Pile-up Peeling}

If pile-up is severe then essentially no method will work~\cite{mcleanLimitationsDecisionBased2018}. If pile-up is only moderate then heuristic methods with low complexity might succeed in identifying the individual pulses. When used in conjunction with pile-up rejection (\S\ref{sec:pile_up_rejection}), fewer pulses need be rejected outright. This is important for real-time applications such as security scanning of baggage and monitoring in advanced nuclear reactors where it is necessary to quickly form an estimate of the underlying distribution~\cite{saxenaDigitalPulseDeconvolution2020a}. 

Pile-up peeling is one such heuristic. It fits the start of a pulse to the data then subtracts the whole fitted pulse. A second pulse can then be fitted to the residual and the process repeated until no more pulses are left. This has been used in~\cite{yangHighThroughputRate2020} to increase throughput.

\subsection{Sparsity and Compressed Sensing}

Imposing no constraints on the amplitudes and arrival times of the pulses leads to an MLE formulation (\S\ref{sec:mle}) while imposing statistical constraints leads to a noisy filtered Poisson process formulation (\S\ref{sec:fpp}). The resulting optimisation problems can be computationally prohibitive.

The popularity of compressed sensing approaches in signal processing~\cite{candesIntroductionCompressiveSampling2008} makes it natural to consider replacing the Poisson constraint on arrival times by a sparsity constraint. In the first instance, the model \eqref{eq:yt} is replaced by the discrete-time model
\begin{equation} \label{eq:discrete}
    \begin{pmatrix}
        y(0) \\ y(1) \\ \vdots
    \end{pmatrix}
    =
    \begin{pmatrix}
        p(0) & p(1) & p(2) & \cdots \\
        0 & p(0) & p(1) & \cdots \\
        \vdots & \vdots & \vdots \\
    \end{pmatrix}
    \begin{pmatrix}
        a(0) \\ a(1) \\ \vdots
    \end{pmatrix}
    +
    \begin{pmatrix}
        w(0) \\ w(1) \\ \vdots
    \end{pmatrix}
\end{equation}
If $a(k) \neq 0$ then a pulse has occurred at time $k$ with amplitude $a(k)$. In the notation of \eqref{eq:yt} this is equivalent to $\tau_i = k$ and $\alpha_i = a(k)$ for some $i$. Clearly, \eqref{eq:discrete} does not perfectly capture pulses arriving at non-integer times, however, one might begin with the naive hope the discrepancy can be absorbed into the noise terms $w(k)$ without much effect.

Being linear, the model \eqref{eq:discrete} lends itself to a least-squares regression. A positivity constraint $a(k) \geq 0$ can be introduced without difficulty because the problem remains convex. Even with this constraint though, the presence of noise means none of the $a(k)$ are likely to be zero even if the data were generated using only a small number of pulses. In an attempt to overcome this, a regularisation term can be added to favour sparse solutions. Writing \eqref{eq:discrete} in vector form as $\mathbf{y} = \mathbf{P}\mathbf{a} + \mathbf{w}$, we can seek the elementwise non-negative vector $\mathbf{a}$ minimising $\| \mathbf{y} - \mathbf{P} \mathbf{a} \|^2 + c\,f(\mathbf{a})$ where $c > 0$ controls the amount of regularisation and $f(\cdot)$ is a real-valued function which is small when its vector argument is sparse.

Choosing the regularisation term $f(\mathbf{a})$ to be the ``0-norm'', meaning it counts the number of non-zero elements of $\mathbf{a}$ (which is not a true norm!), would directly impose the sparsity constraint but lead to a difficult combinatorial optimisation problem. A prevalent idea from compressed sensing is that, for large problems where the matrix $\mathbf{P}$ is sufficiently nice, the 0-norm can be replaced by the 1-norm and the resulting regularised optimisation problem will still find the best sparse solution. The advantage of the 1-norm is that the problem is restored to being convex and thus computationally tractable.

There are three potential shortcomings of this approach.
\begin{enumerate}
    \item A sparsity constraint differs from a Poisson constraint.
    \item The implicit quantisation of pulse arrival times to integer values is a poor fit if pulses have sharp rise times.
    \item The Toeplitz structure of $\mathbf{P}$ may not lend itself to allowing the 1-norm to replace the 0-norm without affecting the optimal solution.
\end{enumerate}

When this sparse pulse-fitting method is applied in practice, it is quite common for a single pulse to be decomposed into two adjacent pulses. This is referred to as the verbosity problem in~\cite{kafaeePileCorrectionSpectroscopicSignals2020}. There, its cause was attributed to the observation that a pulse starting at a non-integer time has a significantly different shape when sampled at integer times than a pulse starting at an integer time; this is the second shortcoming listed above. It is therefore not surprising that non-aligned pulses are better represented using a sum of two pulses than a single pulse. The workaround in~\cite{kafaeePileCorrectionSpectroscopicSignals2020} was to add extra regularisation terms to penalise pulses starting within two units of time from each other.

We offer an additional explanation for the verbosity problem. By continuity, $\alpha_1\, p(t) + \alpha_2\, p(t) = (\alpha_1 + \alpha_2)\, p(t)$ implies $\alpha_1\, p(t-\tau_1) + \alpha_2\,p(t-\tau_2) \approxeq (\alpha_1 + \alpha_2) \, p(t-(\tau_1+\tau_2)/2)$. In particular, two small pulses displaced by a small amount and summed together will look like a single larger pulse almost everywhere except at the sharp rising edge. The bottom left graph in Figure~\ref{fig:pp} does not have a smooth rising edge, despite being a single pulse, and hence it might be better fit using a sum of two or more small pulses than one large one. This will give a better fit to the rising edge without affecting the fit in the tail region. While a sparsity constraint will indeed attempt to reduce the number of pulses used, two pulses will still be favoured over one if the fit is significantly better, as might well be on sharp rising edges corrupted by noise.

The aforementioned paper~\cite{kafaeePileCorrectionSpectroscopicSignals2020} from the nuclear spectroscopy literature might have been unaware of earlier developments in the signal processing literature. Motivated precisely by the second and third shortcomings listed earlier, the continuous basis pursuit methodology was proposed in~\cite{ekanadhamRecoverySparseTranslationInvariant2011}. Rather than use only shifted versions of the pulse $p(t)$ for the rows of $\mathbf{P}$ in \eqref{eq:discrete}, continuous basis pursuit adds other rows that help interpolate between the different possible pulse shapes. A simple example, motivated by a Taylor series, is using shifted versions $p'(t-k)$ of the derivative $p'(t)$ of $p(t)$. The optimisation problem is modified to ensure that if $p(t-k)$ is not present in the decomposition then neither is $p'(t-k)$. An extended version of continuous basis pursuit was developed in~\cite{knudsonInferringSparseRepresentations2014} for fitting neural spike data.

\section{Spectrum Estimation}
\label{sec:spec_est}

This section focuses on the problem of estimating the amplitude distribution $\mathcal{A}$ from the observations $y(t)$ in \eqref{eq:yt} where the $\alpha_i$ are drawn independently from $\mathcal{A}$ and the arrival times $\tau_i$ are Poisson distributed with fixed parameter $\lambda$. This corresponds to the nuclear spectroscopy problem under the assumption that the detector being used consistently produces pulses of the same shape $p(t)$. The extent to which this assumption holds depends on the type of detector being used. Throughout, the Poisson intensity $\lambda$, which is also referred to as the count rate, is assumed to be known. If it is not, it can be estimated. See~\cite{knollRadiationDetectionMeasurement2010} for additional background.

The majority of approaches take a two-step approach: first estimate the amplitudes $\alpha_i$ of the individual pulses then use a histogram of these amplitude estimates as an estimate of the distribution $\mathcal{A}$. Any of the approaches in \S\ref{sec:foundations} can be used to estimate the pulse amplitudes. In particular, pile-up rejection can be employed. Implementations include~\cite{abbeneRealTimeDigital2013,difulvioDigitalPulseProcessing2016,asztalosGeneralPurposePulse2016,leeDevelopmentFPGAbasedDigital2013,saxenaDigitalPulseDeconvolution2020a,yangHighThroughputRate2020,caoFlexibleLowcostFPGAbased2021}.

There are other approaches though. Traditional pile-up correction algorithms take the amplitude distribution estimated in the usual way from the peaks of pulses without (aggressive) pile-up rejection, then apply a transformation to correct for errors caused by pile-up. Indeed, given the amplitude distribution $\mathcal{A}$ and the Poisson intensity $\lambda$, it is possible to calculate the probability of two pulses, one of amplitude $\alpha_1$ and one of amplitude $\alpha_2$, arriving sufficiently close together to be detected as a single pulse of amplitude $\alpha_1 + \alpha_2$. In this way, and by ignoring the possibility of three or more pulses piling up, the transformation from the actual distribution $\mathcal{A}$ to the measured distribution $\mathcal{A}'$ can be determined. Pile-up correction methods take the estimated distribution then invert the mapping $\mathcal{A} \mapsto \mathcal{A}'$ to obtain a corrected distribution~\cite{wielopolskiPredictionPulseheightSpectral1976,barradasAccurateCalculationPileup2006}.

Some more recent methods bypass the need to identify individual pulses altogether, and instead go directly from statistical properties of $y(t)$ to an estimate of $\mathcal{A}$. One class uses the concept of pulse clusters. Given an amplitude threshold $c$, a pulse cluster is an interval $[t_1,t_2]$ such that $y(t) \geq c$ for all $t \in [t_1,t_2]$ and such that the interval cannot be enlarged further. Associated with any such pulse cluster are two parameters, the cluster duration $t_2-t_1$ and cluster area $\int_{t_1}^{t_2} y(t)\,dt$. If there were no pile-up then this would find individual pulses and the area would be directly proportional to the amplitudes $\alpha_i$ whose distribution we seek to estimate\footnote{In nuclear spectroscopy, it is the area under each pulse that is directly proportional to the energy of the incoming photon. If the pulse shape is constant then the area under a pulse and the amplitude of the pulse are proportional to each other. In early designs, it was easier to find a pulse's amplitude than area.}.

Let the $i$th cluster have duration $X_i$ and area $Y_i$. Each pair $(X_i,Y_i)$ can be treated as having been drawn independently from a fixed distribution denoted by $p(X,Y)$. This distribution $p(X,Y)$ is fully determined by the amplitude distribution $\mathcal{A}$ and the Poisson intensity $\lambda$. Therefore, there is a mapping from the distribution $\mathcal{A}$ to the joint distribution $p(X,Y)$. In principle, $\mathcal{A}$ can be estimated by first forming an empirical estimate of $p(X,Y)$ then inverting the mapping $\mathcal{A} \mapsto p(X,Y)$. Approximate methods for achieving this can be found in~\cite{triganoFastDigitalFiltering2015} and the references therein.

Another direct method (\cite{ilheNonparametricEstimationShotnoise2016}) uses the empirical first-order distribution of the observations $y(t)$ to estimate $\mathcal{A}$. In its derivation it neglects the additive noise and works with a (non-noisy) filtered Poisson process (\S\ref{sec:fpp}). Such a process, if considered on the whole real line, is stationary. There is a mapping from $\mathcal{A}$ to the distribution $p(Y_0)$ of $y(0)$. Stationarity means the distribution of $y(0)$ can be estimated empirically from the realisations $y(0), y(1), \cdots$. Applying the inverse of the mapping $\mathcal{A} \mapsto p(Y_0)$ to the empirically estimated distribution of $y(0)$ yields an estimate of $\mathcal{A}$.

Yet another direct method (\cite{mcleanNonParametricDecompoundingPulse2020}) uses a novel statistic. A length $L$ is chosen in advance. Time is broken into non-overlapping intervals $[T_j,T_j+L)$ where $|y(T_j)|$ and $|y(T_j+L)|$ are both smaller than a predetermined threshold. Ideally, each such interval fully contains zero or more pulses; no pulses spill out of the interval. The area $x_j = \int_{T_j}^{T_j+L} y(t)\,dt$ in each interval is the statistic of interest; it corresponds to the total photon energy. The $x_j$ are independent and identically distributed with some distribution $p(X)$ depending on $\mathcal{A}$. Computing the empirical distribution of the $x_j$ then applying the inverse of the mapping $\mathcal{A} \mapsto p(X)$ yields an estimate of $\mathcal{A}$, just like before.

\section{Challenges and Conclusion} \label{sec:conclusion}

In its pure form, the pulse processing problem is to identify each individual pulse given the noisy observations \eqref{eq:yt}. Applying standard estimation and nonlinear filtering techniques generally leads to computationally intractable algorithms and hence heuristics and approximations are highly desirable. The main cause of this computational burden is pulse pile-up. The broad challenge in the field is to develop algorithms that can handle more and more pile-up. There are fundamental limits though; as the distance between two pulses goes to zero, the two pulses become indistinguishable from a single pulse~\cite{mcleanLimitationsDecisionBased2018}. How much of a problem this is depends on the underlying assumptions and objectives. Three typical scenarios are now discussed in turn, corresponding to the three formulations described at the start of \S\ref{sec:foundations}.

\paragraph*{No Statistical Information}
If nothing is known about the pulse amplitudes $\alpha_i$ and arrival times $\tau_i$, the model \eqref{eq:yt} can be interpreted as a nonlinear regression problem (\S\ref{sec:nlr}): find the $\hat\alpha_i$ and $\hat\tau_i$ such that $\hat y(t) = \sum_i \hat\alpha_i p(t - \hat\tau_i)$ best fits the observations $y(t)$. If the definition of best fit is minimising the squared error $\sum_t (\hat y(t) - y(t))^2$, it becomes a nonlinear least-squares problem. Not knowing the number of pulses adds a model-order selection problem on top~\cite{stoicaModelorderSelectionReview2004}. Beyond preprocessing techniques such as pulse shaping (\S\ref{sec:pulse_shaping}), we are unaware of any refinements to the nonlinear least-squares problem specifically for detecting pulses. This motivates the following challenges.

\begin{itemize}
    \item Use properties of the pulse shape $p(t)$ to refine non-linear least-squares techniques.
    \begin{itemize}
        \item First detect clusters of pulses, then fit pulses within each cluster.
        \item Develop pulse-specific model-order selection rules generalising the idea of using a hypothesis test to distinguish two piled-up pulses from a single pulse~\cite{mcleanLimitationsDecisionBased2018}.
    \end{itemize}
    \item Attempt to understand how the locations of the critical points (such as local minima) of the mean-square residue \eqref{eq:msr} vary as noise is introduced, and use this knowledge to develop algorithms for finding the global minimum~\cite{mantonOptimisationGeometry2013} given knowledge of the number of pulses present.
\end{itemize}

\paragraph*{Stochastic Pulse Heights and Arrival Times}
If the pulse amplitudes $\alpha_i$ and arrival times $\tau_i$ are generated from stochastic processes with known statistics, estimating them is a problem in Bayesian statistics. The specific situation where the arrival times are generated by a Poisson process was discussed in \S\ref{sec:fpp}. If the distribution of the $\alpha_i$ is also known, the estimation problem can be treated as a nonlinear regression problem with a regularisation term to account for this prior information, and the challenges listed above remain applicable. The situation where the amplitude distribution is unknown is considered below.

\paragraph*{Estimation of Amplitude Distribution}

The pulse processing problem most studied is that of estimating the distribution $\mathcal{A}$ of the pulse amplitudes $\alpha_i$ under the assumption that each $\alpha_i$ is drawn independently at random from $\mathcal{A}$. This is the core challenge in nuclear spectroscopy, for example, and it has two advantages over the more general problem. These advantages facilitate the derivation of tractable solutions. First, it is allowable to reject piled-up pulses (\S\ref{sec:pile_up_rejection}), affording a trade-off between tractability and efficiency. Second, statistical techniques can be applied, either from the start, or as post-processing refinements (\S\ref{sec:spec_est}). The main challenge then is improving the efficiency of these algorithms, meaning decreasing the number of pulses needed to form an accurate estimate of the amplitude distribution. Potential future directions include the following.

\begin{itemize}
    \item Rather than identify individual pulses, use the statistics of the observed noisy filtered Poisson process to determine the pulse amplitude distribution directly. (Approximate methods exist but they do not use all the available statistical information.)
    \item Develop ``soft'' estimates giving different possible decompositions of a signal $y(t)$ into a sum of pulses along with a reliability score for each one.
    \item Develop an iterative refinement procedure whereby the current estimate of the amplitude distribution can be used to help separate piled-up pulses, and these separated pulses can then be used to improve the estimate of the amplitude distribution.
\end{itemize}

An over-arching challenge is establishing rigorous mathematical performance bounds. This would aid in assessing the veracity of our belief that there is substantial potential for improving pulse processing algorithms.

\section*{Acknowledgement}

We thank the anonymous reviewers for excellent suggestions which enhanced the quality of this paper.

\bibliographystyle{elsarticle-num}
\bibliography{pulse}
\end{document}